\documentclass[useAMS,usenatbib]{mn2e}

\usepackage[T1]{fontenc}
\usepackage{pslatex}
\usepackage{amsmath}
\usepackage{amsfonts}
\usepackage{color}
\usepackage{url}
\usepackage{graphicx}
\usepackage{subfigure}
\usepackage{amssymb}
\usepackage{textcomp}
\usepackage{soul}
\usepackage{booktabs}
\usepackage{array}
\usepackage{hyperref}
 \graphicspath{{images/}}

{\newif\ifnotend
\notendtrue
\def\veclist{ABCDEFGHIJKLMNOPQRSTUVWXYZabcdefghijklmnopqrstuvwxyz.}
\def\top#1#2.{#1}
\def\tail#1#2.{#2.}
\loop\expandafter\xdef\csname v\expandafter\top\veclist\endcsname%
{{\noexpand\bf\expandafter\top\veclist}}
\edef\veclist{\expandafter\tail\veclist}
\if\veclist.\notendfalse\fi\ifnotend\repeat}
\def\pa{\partial}

\mathchardef\mhyphen="2D

\title[Nuclear Spirals]{Nuclear spirals in the inner Milky Way}
\author[Ridley, Sormani, Tre{\ss}, Magorrian, Klessen]{Matthew Ridley$^{1,\star}$, Mattia C. Sormani$^{2,\star}$, Robin G. Tre{\ss}$^2$, John Magorrian$^1$ \newauthor and Ralf S. Klessen$^{2,3}$ \\
$^1$Rudolf Peierls Centre for Theoretical Physics, 1 Keble Road, Oxford OX1 3NP, UK\\
$^2$Universit\"{a}t Heidelberg, Zentrum f\"{u}r Astronomie, Institut f\"{u}r theoretische Astrophysik, Albert-Ueberle-Str. 2, 69120 Heidelberg, Germany \\
$^3$Universit\"at Heidelberg, Interdiszipli\"ares Zentrum f\"ur Wissenschaftliches Rechnen, Im Neuenheimer Feld 205, 69120 Heidelberg, Germany \\
$^\star$These two authors contributed equally to this paper
}

\begin{document}
\hyphenation{kruijs-sen}

\date{}

\def\p{\partial}
\def\Omegap{\Omega_{\rm p}}

\newcommand{\di}{\mathrm{d}}
\newcommand{\bfx}{\mathbf{x}}
\newcommand{\bfe}{\mathbf{e}}
\newcommand{\vlos}{\mathrm{v}_{\rm los}}
\newcommand{\Tspin}{T_{\rm s}}
\newcommand{\Tb}{T_{\rm b}}
\newcommand{\degree}{\ensuremath{^\circ}}
\newcommand{\Th}{T_{\rm h}}
\newcommand{\Tc}{T_{\rm c}}
\newcommand{\bfr}{\mathbf{r}}
\newcommand{\bfv}{\mathbf{v}}
\newcommand{\pc}{\,{\rm pc}}
\newcommand{\kpc}{\,{\rm kpc}}
\newcommand{\Myr}{\,{\rm Myr}}
\newcommand{\Gyr}{\,{\rm Gyr}}
\newcommand{\kms}{\,{\rm km\, s^{-1}}}
\newcommand{\de}[2]{\frac{\partial #1}{\partial {#2}}}
\newcommand{\cs}{c_{\rm s}}
\newcommand{\rb}{r_{\rm b}}
\newcommand{\rqu}{r_{\rm q}}
\newcommand{\nuP}{\nu_{\rm P}}
\newcommand{\thetaobs}{\theta_{\rm obs}}
\newcommand{\hatn}{\hat{\textbf{n}}}
\newcommand{\hatx}{\hat{\textbf{x}}}
\newcommand{\haty}{\hat{\textbf{y}}}
\newcommand{\hatz}{\hat{\textbf{z}}}
\newcommand{\hatX}{\hat{\textbf{X}}}
\newcommand{\hatY}{\hat{\textbf{Y}}}
\newcommand{\hatZ}{\hat{\textbf{Z}}}
\newcommand{\hatN}{\hat{\textbf{N}}}

\maketitle

\begin{abstract}
We use hydrodynamical simulations to construct a new coherent picture for the gas flow in the Central Molecular Zone (CMZ), the region of our Galaxy within $R\lesssim500\pc$. We relate connected structures observed in $(l,b,v)$ data cubes of molecular tracers to nuclear spiral arms. These arise naturally in hydrodynamical simulations of barred galaxies, and are similar to those that can be seen in external galaxies such as NGC4303 or NGC1097. We discuss a face-on view of the CMZ including the position of several prominent molecular clouds, such as Sgr B2, the $20\kms$ and $50\kms$ clouds, the polar arc, Bania Clump 2 and Sgr C. Our model is also consistent with the larger scale gas flow, up to $R\simeq3\kpc$, thus providing a consistent picture of the entire Galactic bar region.
\end{abstract}


\section{Introduction}
\label{sec:intro}

Since the earliest radio surveys it has been known that the interstellar medium (ISM) in the Central Molecular Zone (CMZ), the innermost $~500\pc$ of the Milky Way (MW), has a particularly rich and complex structure. The kinematics show significant departures from circular motion, the molecular gas is strongly concentrated within the central degree, and emission is highly asymmetric about the Galactic Centre (GC) \citep[e.g.][]{Bally1988}. This region contains some 10$\%$ of the total molecular gas of the Milky Way ($\sim 10^7\,M_\odot$, \citealt{Ferriere2007}), and the physical conditions within the CMZ are far more extreme than in the Solar neighbourhood and outer regions of the Galaxy. Despite the extremely high column densities and pressures, star formation in the CMZ seems to be suppressed by at least an order of magnitude compared to what is predicted by our current understanding of star formation \citep[e.g.][]{Kennicutt98,Longmore+2013,Kruijssen+14b,Battersby+2016}. A proper understanding of the structure and dynamics of gas in the CMZ may help reconcile this contradiction, and give us insight into the mechanics of star formation.

High resolution molecular line studies of the CMZ have revealed coherent features spanning the width of the $(l,b,v)$ data cubes in several different species, including CO, CS, NH$_3$ and HNCO \citep{Nagayama+2007,Oka+2007,Henshaw+16a}. The origin of these features has been of interest for several decades, and proposed explanations include two spiral arms \citep{Sofue95,Sawada+2004}, a twisted elliptical orbit \citep{Molinari+2011}, or an open stream \citep{Kruijssen+2015}.

The dynamics of gas in the CMZ is an important topic in itself. Since it is now well established that the MW contains a bar, as confirmed by direct photometric evidence \citep{Blitz1991}, it is clear that the dynamics must be understood in the context of gas flowing in a barred potential. In the interpretation of \cite{Binney++1991}, later refined by \cite{SBM2015a} \citep[hereafter SBM15a; see also][]{JenkinsBinney94,RFC2008,Li++2016}, gas in the CMZ follows $x_2$ orbits. These are a family of closed orbits weakly elongated in the direction perpendicular to the bar \citep{Contopoulos1980}. Hence the CMZ is sometimes referred to as the $x_2$ disc. The CMZ is fed gas by shocks, which efficiently bring material inwards onto the $x_2$ orbits from the outer parts ($R\simeq2-3\kpc$).

\begin{figure}
\includegraphics[width=84mm]{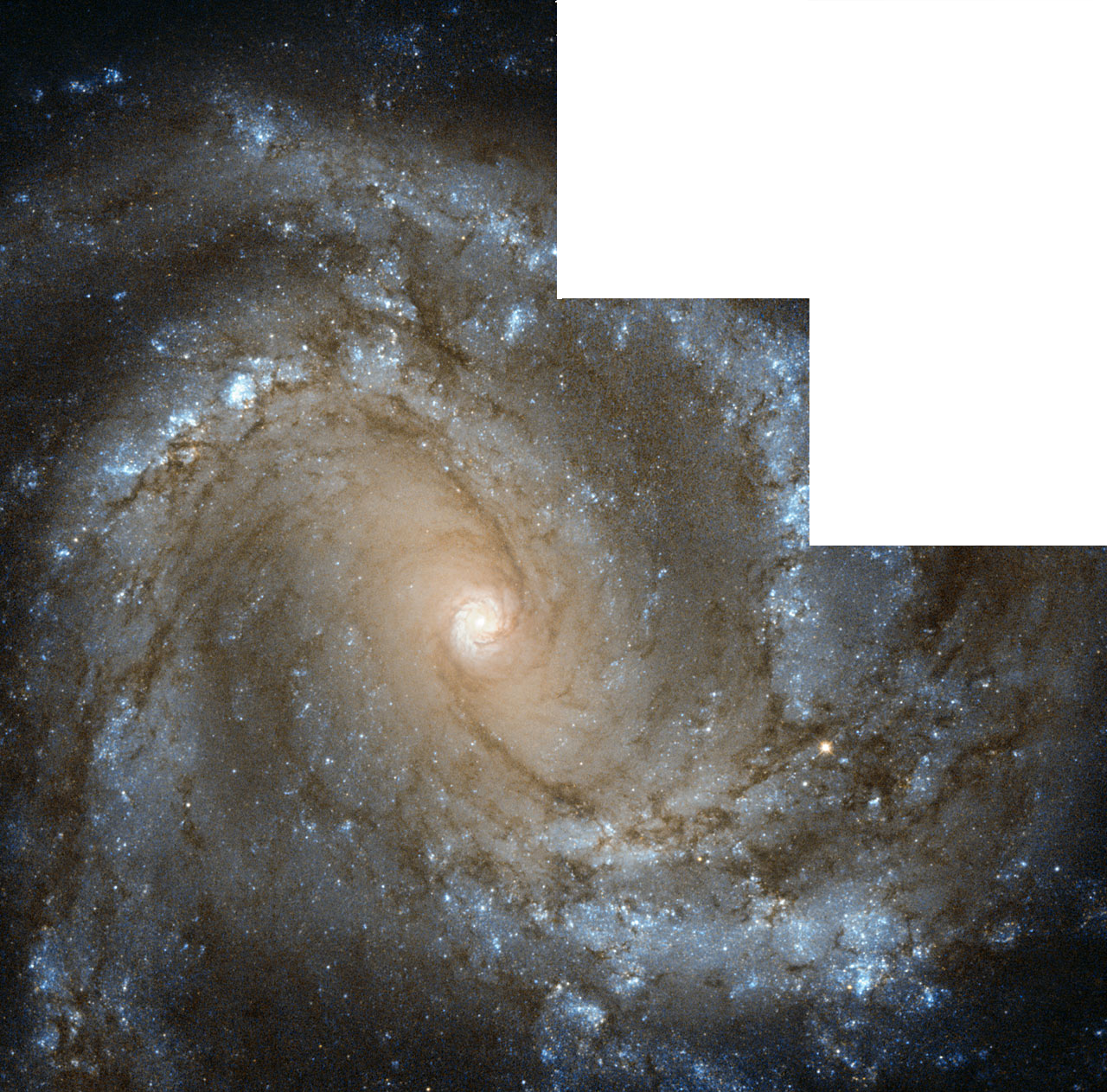}
\caption{
An image of the barred spiral Galaxy NGC4303, also known as Messier 61, captured by the Hubble Space Telescope Wide Field Camera 2. Nuclear spirals fed by straight dust lanes are clearly visible in the core. Credit: ESA/Hubble \& NASA. Acknowledgements: G. Chapdelaine, L. Limatola, and R. Gendler. \url{https://www.spacetelescope.org/images/potw1324a/}.
}
\label{fig:external}
\end{figure}

Our goal in this paper is to demonstrate that the distribution and kinematics of gas in the CMZ can be well modelled by an elongated disc of gas containing two nuclear spiral arms. Nuclear spirals are commonly observed in external barred spiral galaxies \citep[e.g.][]{Schinnerer+2002,Martini+2003c, Martini+2003b,Kuno+2008,DeVenFathi2010}, see Fig. \ref{fig:external} for a striking example. It would therefore be natural to also find them in the centre of our Galaxy. They arise commonly in hydrodynamic simulations in barred potentials \cite[e.g.][]{AnnThakur2005,Li+2015}, thus they are automatically consistent with the large scale flow in and around the bar. If indeed nuclear spirals are present in the centre of the MW, the density enhancement along the spiral arms would result in ridges of emission spanning the width of the CMZ similar to those observed in the $(l,b,v)$ data cubes. The compression produced by the spiral shocks would provide a natural mechanism for the formation of complex molecular species.

The presence of nuclear spiral arms at the centre of the MW has been discussed for some time. \cite{Sofue95} used a purely kinematic spiral arm model to deconvolve the $(l,v)$ distribution of emission in the CMZ to produce a face-on map. \cite{Sawada+2004} compared emission and absorption maps to derive distances of clouds along lines-of-sight without the use of kinematic models, and the resulting face-on map was consistent with the presence of two arms of gas in the central region. However, neither of these was a dynamical model derived from physical principles. Our aim here is to improve on previous work by producing a fully dynamic model of the CMZ consistent with the gas flow in the entire central region of the Galaxy. As we describe below, our picture differs in several significant ways from previous works and corrects some inconsistencies of the spiral arm interpretation of \cite{Sofue95} and \cite{Sawada+2004} as discussed in \cite{Henshaw+16a}.

This paper is structured as follows. In Section \ref{sec:data} we discuss the data used in this paper, and highlight particular features of interest. In Section \ref{sec:methods} we describe the numerical scheme and the potential used to run the hydrodynamical simulation. The results of the model are described in Section \ref{sec:results}, and we discuss the successes and shortcomings of the model in Section \ref{sec:discuss}. Finally in Section \ref{sec:conc} we summarise our conclusions.

\section{Data}
\label{sec:data}

Here we briefly describe some characteristics of the observations. This section should serve as a reference for the remainder of the paper.

We focus on NH$_3$ (1,1) emission in the region $-1^\circ \leq l \leq 3.5^\circ$, using data from the H$_2$O Southern Galactic Plane Survey (HOPS; \citealt{Walsh+2011}, \citealt{Purcell+2012}). The data have a spatial resolution of $\sim$ 60 arcsec and a spectral resolution of 2$\,\mathrm{km}\,\mathrm{s}^{-1}$, and have been fit using the Semi-automated multi-COmponent Universal Spectral-line fitting Engine (\textsc{scouse} \footnote{https://github.com/jdhenshaw/SCOUSE}; \citealt{Henshaw+16a}). These data and the fits are discussed in detail in \cite{Longmore2017}.

Although NH$_3$ shows some signs of self-absorption in the CMZ, the distribution and kinematics of emission show the same large scale features as observed in CO \citep{Rodriguezetal2006}, HNCO \citep{Henshaw+16a} and C\textsc{ii} \citep{Langer2017}. In this paper we focus mainly on coherent features in the ($l,b,v$) data cube, and not on the precise details of the emission.

\begin{figure}
\includegraphics[width=84mm]{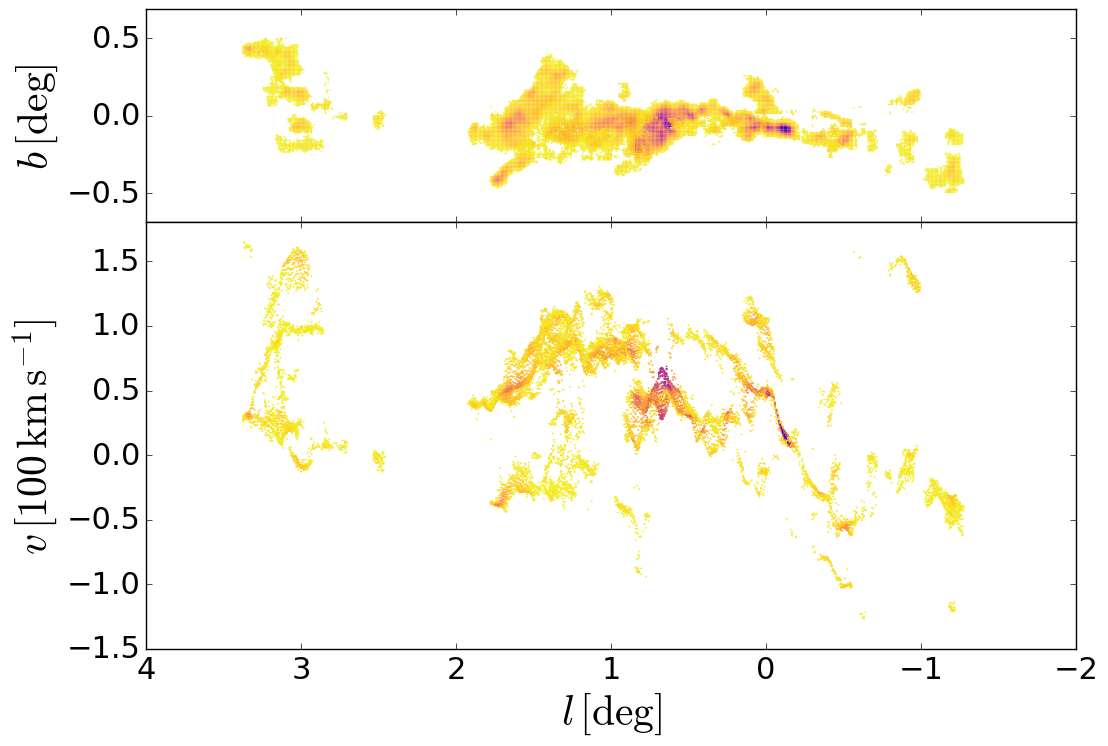}
\caption{NH$_3$ $(1-1)$ data from the HOPS survey \protect\citep{Purcell+2012}, fit with the \textsc{scouse} algorithm \protect\citep{Henshaw+16a}. Top panel: The spatial distribution of emission in ($l,b$). Bottom panel: The position velocity distribution in $(l,v)$. Each point represents a spectral component as determined by the \textsc{scouse} algorithm, coloured by brightness temperature (arbitrary units).}
\label{fig:henshaw}
\end{figure}

\begin{figure}
\includegraphics[width=84mm]{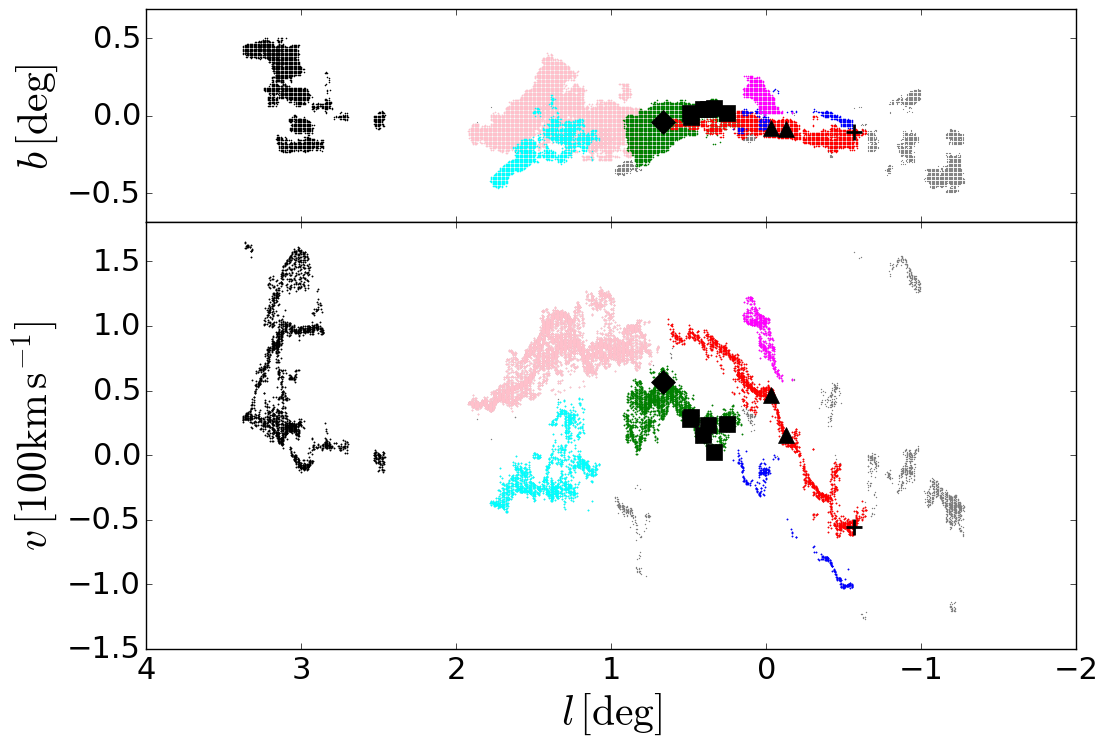}
\caption{The data of Fig. \ref{fig:henshaw}, with features of interest highlighted in various colors. \textbf{Blue points}: Arm \textsc{i}. \textbf{Red points}: Arm \textsc{ii}. \textbf{Magenta points}: Arm \textsc{iii}, also known as the ``polar arc''. \textbf{Green points and black diamond}: Sgr B2 cloud complex and the dust ridge. \textbf{Pink points}: 1.3$^\circ$ cloud complex. \textbf{Black points}: Clump 2 of \protect\cite{StarkBania1986}.  \textbf{Cyan points}: Cloud complex M of \protect\cite{Rodriguezetal2006}. \textbf{Black triangles}: the 20 and 50$\kms$ clouds. \textbf{Black plus}: Sgr C. \textbf{Black squares:} From left to right dust ridge clouds B to F and G0.253+0.016, also known as ``the Brick''.}
\label{fig:features}
\end{figure}

Fig. \ref{fig:henshaw} shows the $(l,v)$ and $(l,b)$ projections of the \textsc{scouse} fits to the data. We note several structures of particular interest, highlighted in Fig. \ref{fig:features}:
\begin{enumerate}
    \item Blue points: Arm \textsc{i}, an extended ridge of emission running from $l=-0.7^{\circ}$, $v = 100\kms$ to $l=0.7^{\circ}$, $v~=~-~50~\kms$.
    \item Red points: Arm \textsc{ii}, another ridge of emission running parallel to Arm \textsc{i} from $l=-0.7^{\circ}$, $v = 70\kms$ to $l=0.7^{\circ}$, $v~=~-150\kms$. \cite{Sofue95} identified Arms \textsc{i} and \textsc{ii} as two spiral arms within the CMZ, however \cite{Kruijssen+2015} proposed that they are the projection of gas clouds on an open ballistic orbit.
    \item Magenta points: Arm \textsc{iii}, also known as the ``polar arc''. A high velocity feature at a large inclination, suggested as a continuation of Arm \textsc{ii} by \cite{Sofue95}
    \item Green points: Sgr B2 cloud complex and the dust ridge. The Sgr B2 cloud is a well studied molecular cloud complex around $(l,b,v) \simeq (0.7^\circ, 0.1^\circ, 10-70\kms)$. An unusually high number of independent velocity components are present at each $(l,b)$ pixel \citep{Henshaw+16a}, and the velocity dispersions are also unusually broad. Suggested explanation for this complex velocity structure include a shell-like arrangement produced by supernovae or the result of a collision between two molecular clouds \citep[e.g.][and references therein]{Henshaw+16a}. Between Sgr B2 and G0.253+0.016 (the ``Brick'', rightmost black square) lies a prominent ridge of dust emission containing several molecular clouds. See section \ref{sec:features} for a more in depth discussion.
    \item Pink points: 1.3$^\circ$ cloud complex, a huge molecular cloud structure suggested by \cite{Rodriguezetal2006} as the site of accretion onto the CMZ.
    \item Black points: Clump 2, a molecular cloud complex at ($l,b,v) \simeq (3^\circ, 0.2^\circ, 20-150\kms)$. \cite{StarkBania1986} suggested that this is the signature of a dust lane or spiral arm due to the complex internal velocity structure.
    \item Cyan points: Cloud M of \cite{Rodriguezetal2006}, possibly associated with the far side dust lane.
  \end{enumerate}

Also shown in the same figure are locations of some prominent molecular clouds. The black triangles are the 20$\kms$ and 50$\kms$ clouds, two bright clouds located near to Sgr A* in projection, believed to be connected to Arm \textsc{ii}. The black plus is Sgr C, a star forming molecular cloud complex at the tip of Arm \textsc{ii}, multiple line-of-sight velocity components. For a more complete discussion, see for example \cite{Rodriguezetal2006,Henshaw+16a}.

\section{Methods}
\label{sec:methods}

\subsection{Numerical scheme}
\label{sec:hydro}

We use the same numerical scheme as in \cite{SBM2015a,SBM2015b,SBM2015c}. Here we briefly summarise the main points, and refer the reader to the above references for further details.

In our simulations, we assume that the gas is a two-dimensional inviscid isothermal fluid governed by the Euler equations. The gas is non-selfgravitating and flows in an externally imposed barred potential that rotates with constant pattern speed $\Omega_{\rm p}$. The potential is described in detail in Section \ref{sec:potential}.

The CMZ has a thickness of 15-30$\,$pc and a radius of $\sim$200$\,$pc \citep{Ferriere2007}. As the vertical extent of the gas is so much smaller that the radial extent, our 2D model should be able to capture the important points of the gas flow.

An additional term is introduced in the continuity equation to implement the recycling law of \cite{Athan92b}. The recycling law was originally meant to take into account the effects of star formation and stellar mass loss in a simple way. In practice, the only effect of the recycling law is to prevent too much gas from accumulating in the very centre and to replace gas lost at the boundary due to the outflow boundary conditions. It does not affect the morphology of the results, so our results do not change if we disable the recycling law. The dynamical equations in an inertial frame are
\begin{equation} \begin{split}
	\pa_t \rho + \nabla \cdot \left( \rho \bfv \right) & = \alpha_{\rm r} (\rho_0^2 - \rho^2),  \\
	\pa_t \bfv + \left( \bfv \cdot \nabla \right) \bfv & = - \frac{\nabla P}{\rho} -\nabla \Phi, \\
	P & = \cs^2 \rho,\,
\end{split} \end{equation}
where $\rho$ is the surface density of the gas, $P$ is the pressure, $\Phi$ is the gravitational potential, $\bfv$ is the velocity, $\cs$ is the sound speed, $\alpha_{\rm r}$ is a constant representing the efficiency of the recycling law and $\rho_0$ is the initial surface density. We adopt recycling efficiency $\alpha_{\rm r}~=~0.3 \, M_\odot \pc^{-2}\Gyr^{-1}$ and initial density $\rho_0 = 1 \, M_\odot\pc^{-2}$ \citep{Athan92b}.

In our simulations the gas is assumed to be isothermal with an effective sound speed of $\cs=10\kms$. This is a phenomenological sound speed that is meant to take into account the effects of small-scale turbulence \citep[e.g.][]{Roberts1969,Cowie1980}; the ``temperature'' in our isothermal assumption is therefore related to the velocity dispersion of clouds rather than a microscopic temperature.  The average energy content of the gas, and hence the effective temperature, is assumed to be held constant by an energy balance between heating and cooling processes. The observed velocity dispersion of the interstellar medium seems to support this hypothesis \citep[e.g.][]{DickeyLockman1990,Walter2008,Leroy2009}. Therefore any heating due to compression, for example at a shock, is instantaneously radiated away to restore the initial temperature.
SBM15a found that the choice $\cs\gtrsim10\kms$ makes the location of the transition from the $x_1$ to $x_2$ families of orbits consistent with observations.

We use a grid-based, Eulerian code based on the second-order flux-splitting scheme developed by \cite{vanAlbada+82} and later used by \cite{Athan92b}, \cite{weinersellwood1999} and others to study gas dynamics in barred potentials.

Our grid is $1750 \times 1750$ on a side. The resolution is $\di x~=~5 \pc$, so the total simulated area is a square with side $8.75\kpc$. We start with gas in equilibrium on circular orbits in an axisymmetrised potential and turn on the non-axisymmetric part of the potential gradually during the first $150\,\rm Myr$ to avoid transients. We use outflow boundary conditions: gas can freely escape the simulated region, after which it is lost forever. The potential well is sufficiently deep, however, to prevent excessive quantities of material from escaping.

To project material to the $(l,v)$ plane, we assume an angle $\phi~=~20^{\circ}$ between the Sun-GC line and the bar major axis (the $x$-axis of the simulation), consistent with current estimates \citep[e.g.][]{BHG16}, and a Sun-GC distance of $R_0=8\kpc$.

\subsection{The potential}
\label{sec:potential}

We use a realistic MW potential that is the sum of 4 components: bar, bulge, disc, halo. As a starting point we use the best fitting potential of \cite{McMillan2017}. This is an axisymmetric model, and therefore cannot accurately represent the central region of the Galaxy, which is dominated by a rotating stellar bar. As we focus on this region, and in particular on the effects of the bar, we perform the following modifications:
\begin{enumerate}
\item We add a barred component. This is chosen such that its quadrupole fits the best fitting quadrupole in \cite{SBM2015c}.
\item We modify the inner bulge density profile to make it more centrally concentrated, so that the potential has an Inner Lindblad Resonance (ILR) which is required for the $x_2$ orbit family to exist as these are believed to provide the backbone for the CMZ.
\item We adjust the parameters of the other components to compensate for the introduction of the barred component. This means that we slightly decrease the mass of the disc and the halo in order to compensate for the extra mass introduced by the bar, in such a way that the circular velocity at the Sun position remains approximately constant.
\end{enumerate}

We can expand the potential in the plane of the Galaxy in multipoles as follows:
\begin{equation}
\Phi(R,\phi) = \Phi_0(R) + \sum_{m=1}^{\infty} \Phi_m(R) \cos\left(m \phi + \phi_m\right),
\end{equation}
where $\phi_m$ are constants and $\{R,\phi,z\}$ denote planar polar coordinates.

\begin{figure}
\includegraphics[width=84mm]{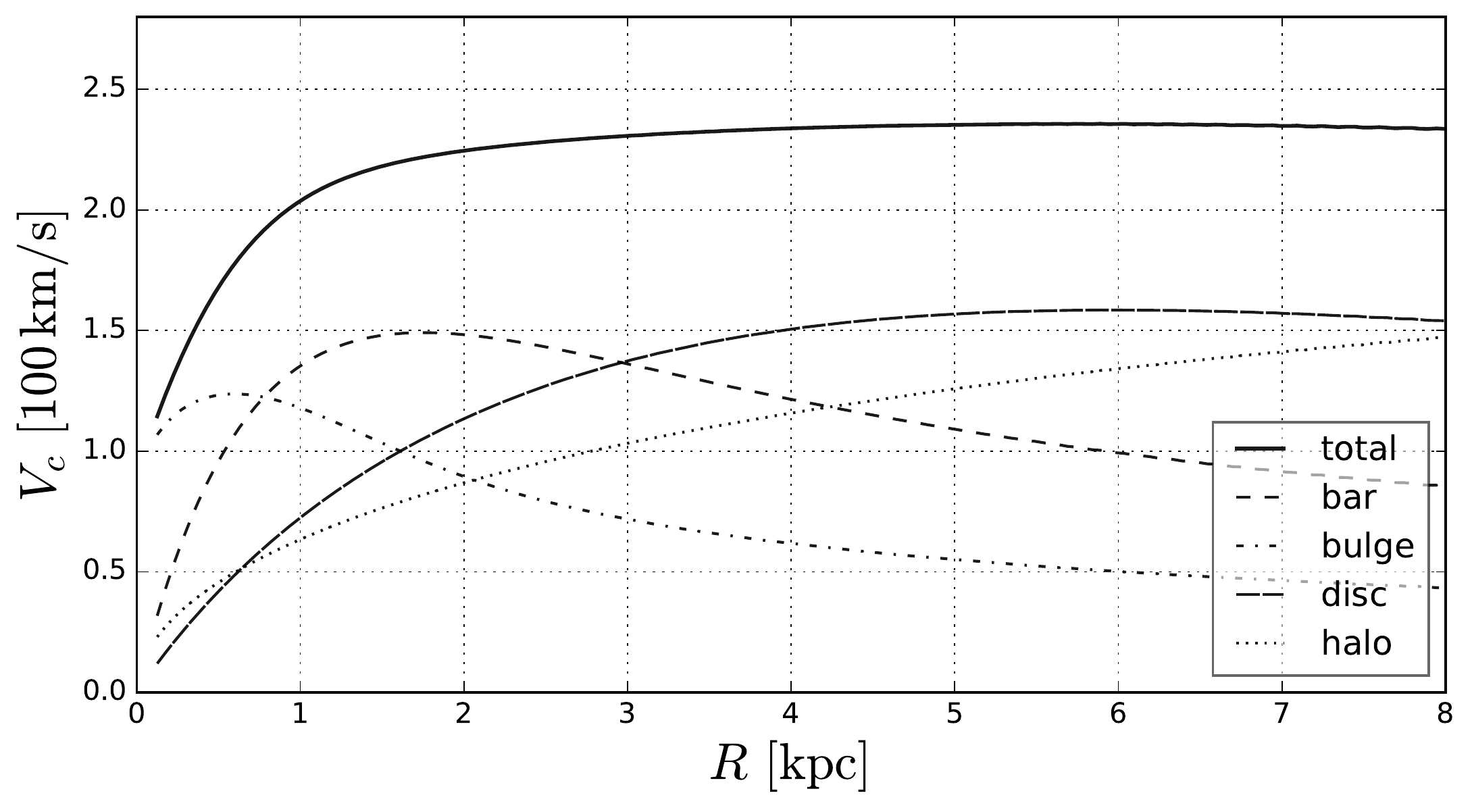}
\caption{The full curve shows the circular velocity curve for the potential used in this paper. The separate contributions from bulge, disc and halo are shown with dot-dashed, dashed and dotted lines respectively.}
\label{fig:vc}
\end{figure}

Fig. \ref{fig:vc} shows the circular velocity curve of the potential and the contributions from each component separately. Note that our definition of ``circular velocity curve'' is based on the axisymmetric part of the potential:
\begin{equation}
V_c(R) \equiv \sqrt{ R \frac{\di \Phi_0}{\di R} }.
\end{equation}
Since the gas undergoes strong non-circular motions in the region dominated by the bar, which for our Galaxy corresponds approximately to the region within Galactocentric radius $R = 3 \kpc$ \citep[e.g.][]{BM}, the ``circular velocity speed'' can be significantly different from the speed of the gas obtained in simulations, or observed in the Galaxy, at the same radii \cite[e.g.][SBM15a]{Binney++1991}.

\begin{figure}
\includegraphics[width=84mm]{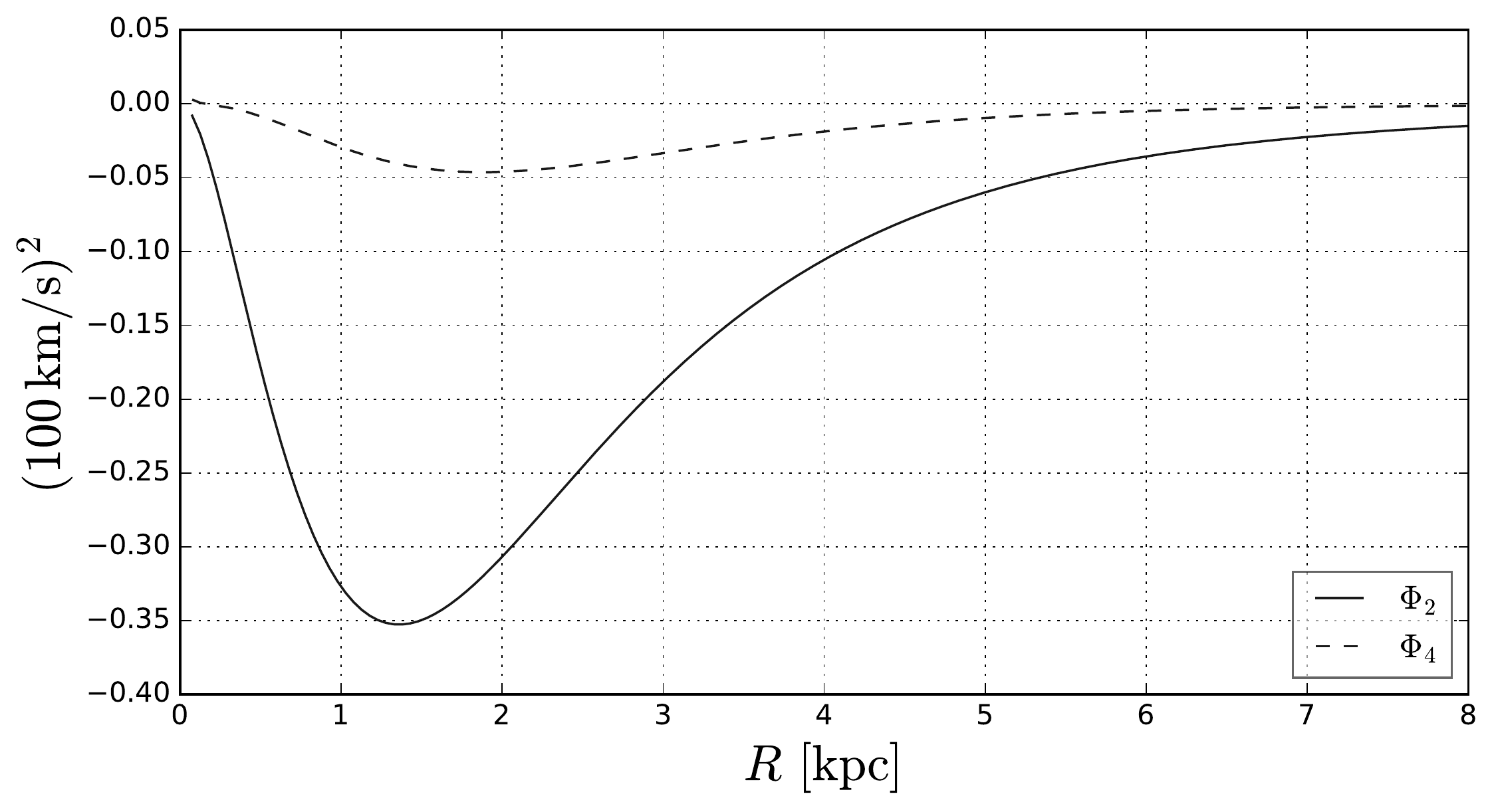}
\caption{The quadrupole $\Phi_2$ and octupole $\Phi_4$ of the potential used in this paper.}
\label{fig:multipoles}
\end{figure}

Fig. \ref{fig:multipoles} shows the quadrupole and the octupole of the potential used in this paper. These are generated by the bar, which is the only non-axisymmetric component in our potential.

The details of each component of the potential are as follows.

\subparagraph{Bulge.} This component is generated by the following density distribution:
\begin{equation}
\rho_{\rm b} = \frac{ \rho_{{\rm b}0} }{(a/a_0)^\alpha} \exp\left[ - \left( a/a_{\rm cut}\right)^2 \right]
\end{equation}
where
\begin{equation}
a = \sqrt{x^2 + y^2 + \frac{z^2}{q_{\rm b}^2}}.
\end{equation}
We use the following values for the parameters: $\alpha=1.7$, $\rho_{{\rm b}0}~=~0.8   M_\odot \pc^{-3}$, $a_{\rm cut}=1.0\kpc$, $q_{\rm b}=0.5$, and without loss of generality we arbitrarily set $a_0=1.0\kpc$. The value $\alpha=1.7$ for the inner slope ($R \lesssim 500 \pc$) of the density distribution is consistent with near-infrared photometry \citep[e.g.][]{BissantzGerhard2002,Launhardt+2002}. We cut the bulge at $a_{\rm cut}=1.0\kpc$ as we assume that beyond this radius the bar dominates over the bulge.

\subparagraph{Bar.} The density of the bar is taken to be:
\begin{equation}
\rho_{\rm B} = \rho_{{\rm B}0}  \exp\left( - a/a_{\rm B} \right),
\end{equation}
where
\begin{equation}
a=\sqrt{x^2 + \frac{y^2+z^2}{q_{\rm B}^2} },
\end{equation}
with the following values for the parameters: $\rho_{{\rm B}0}=5  M_\odot \pc^{-3}$, $a_{\rm B}=0.75\kpc$ and $q_{\rm B}=0.5$. The bar is also assumed to rotate with constant pattern speed of $\Omega_{\rm p} = 40\kms \kpc^{-1}$. This places the Inner Lindblad Resonance at $R_{\rm ILR}=1.2 \kpc$. The form of the bar density distribution is chosen to be exponential as infrared photometry has found the Milky Way bar density profile to be roughly exponential \citep[][]{WeggGerhard2013}. The parameters, including the pattern speed, have been chosen such that the quadrupole of the bar matches the best fit quadrupole of \cite{SBM2015c}.

\subparagraph{Disc.} We assume that the disc is the sum of a thick and a thin disc \citep{GilmoreReid1983}. The density distribution is:
\begin{equation}
\rho_{\rm d} = \frac{\Sigma_1}{2 z_1} \exp \left( -\frac{|z|}{z_1} - \frac{R}{R_{{\rm d} 1}} \right) + \frac{\Sigma_2}{2 z_2} \exp \left( -\frac{|z|}{z_2} - \frac{R}{R_{{\rm d} 2}} \right),
\end{equation}
where $\Sigma_1~=~850 M_\odot \kpc^{-2}$, $R_{\rm{d}1} = 2.5\kpc$, $z_1=0.3\kpc$, $\Sigma_2~=~174 M_\odot \kpc^{-2}$, $R_{\rm{d}2}=3.02\kpc$, $z_2=0.9\kpc$. These parameters are slight modifications of the parameters of \cite{McMillan2017}.

\subparagraph{Halo.} This is a simple \cite{NFW96} profile. The density distribution is:
\begin{equation}
\rho_{\rm h} = \frac{\rho_{{\rm h}0}}{x (1 + x)^2}
\end{equation}
where $x = r/r_h$, $\rho_{\rm{h}0}=0.00811 M_\odot \pc^{-3}$, and $r_h = 19.6\kpc$. This is a slight modification of the best fit model of \cite{McMillan2017}.

\begin{figure}
\includegraphics[width=84mm]{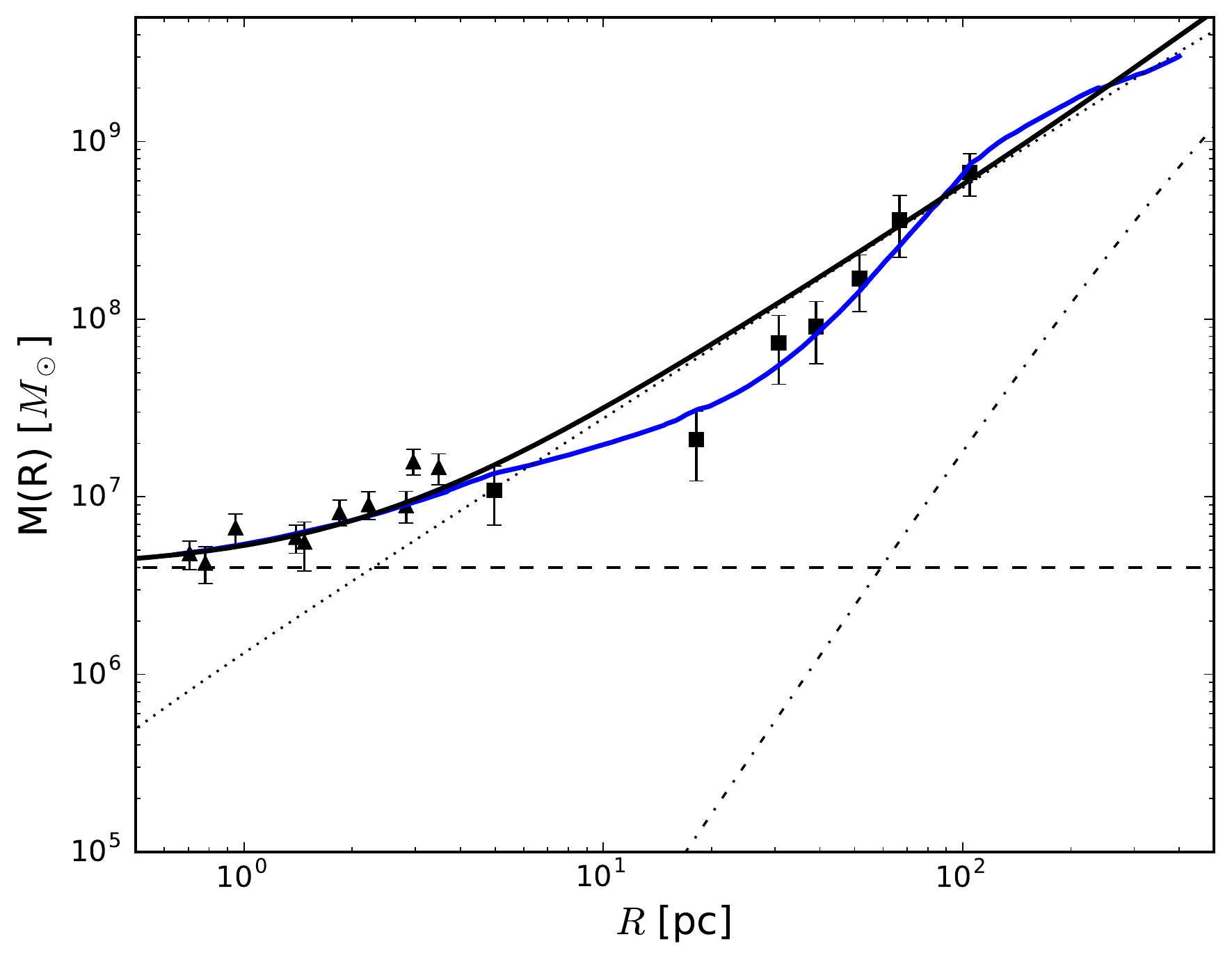}
\caption{Enclosed mass within spheres of radius $R$ for our model, compared to dynamic and photometric estimates for the Milky Way. \textbf{Dashed line:} A central black hole of mass $4 \times 10^{6} M_\odot$. \textbf{Dotted line:} The bulge. \textbf{Dash-dotted line:} The bar. \textbf{Solid black line:} The sum of all three components. \textbf{Solid triangles:} \protect\cite{McGinn89}. \textbf{Solid squares:}  \protect\cite{Lindqvist92}. \textbf{Solid blue line:} the total enclosed mass form \protect\cite{Launhardt+2002}, see their figure 14.}
\label{fig:mass_inc}
\end{figure}

Figure \ref{fig:mass_inc} compares our stellar mass model of the central region of the MW with results of \cite{Launhardt+2002} derived from infrared photometry. We find that our model agrees well with the data. We do not explicitly include the potential due to the central black hole Sgr A* in our simulations as it has no effect on the gas flow at the scales we consider here.

\section{Gas dynamics}
\label{sec:results}

\subsection{Large-scale gas flow} \label{sec:largerscale}

Figure \ref{fig:rho} shows the surface density of the hydro simulation at $t~=~280\Myr$. At this time the gas flow has reached an approximate steady state. The large-scale (\textasciitilde few kpc) gas flow in and around the bar can be described along the lines detailed in \cite{SBM2015a,SBM2015b,SBM2015c}. We briefly recap the main points here, and refer the reader to the above papers for further details. Far from the centre, outside $R \sim 1$kpc, the gas closely follows closed orbits belonging to the $x_1$ family, which are highly elongated along the major axis of the bar. While approximately following the $x_1$ orbits, the gas also slowly drifts inwards until, close to the Inner Lindblad Resonance, the ``cusped'' $x_1$ orbit is reached. At this point, the gas cannot continue to follow the $x_1$ orbits because they become self-intersecting. The gas piles up at these cusps, and two almost straight shocks form from gas plunging almost radially from the tips of the $x_1$ orbits towards the CMZ, where it settles on the less-extended, more circular $x_2$ orbits, forming a central disc.

Thus, closed ballistic orbits are the key to understand the gas flow. Note that the word \emph{closed} (as opposed to \emph{open} ballistic orbits) is crucial here. As noted by \cite{Binney++1991}, when clouds of gas are released into a potential, the clouds will in general shear and eventually settle onto closed orbits, as collisions tend to dissipate the excess energy of large excursions around these closed orbits (most open ballistic orbits can be interpreted as excursions around a ``parent'' closed ballistic orbit, which are $x_1$ and $x_2$ orbits in this case). Thus, while pressure forces are important in collimating gas onto closed orbits, they are much less significant during the long periods in which the gas drifts from one closed orbit to the next. In this latter regime, pressure forces are however responsible for the librations that generate the spiral arms as kinematic density waves \citep[see][and the next section]{SBM2015b}. This is why, away from transition between different orbit families, closed ballistic orbits provide a good approximation to the gas flow despite the presence of the pressure term in the equations of motion (see figure 5 in SMB15a).

Many important observational features in the region $| l | \leq 30 \degree$ (e.g. the 3kpc arm, the 135$\kms$ arm, the terminal velocity curve and others; see for example \citealt{Fux1999,SBM2015c,Li++2016}), which approximately corresponds to $R\lesssim3\kpc$, can be explained by a gas flow such as the one described above. The main point of interest for this paper is that the CMZ is embedded in a consistent gas flow that can reproduce features of the whole barred region, and not just of the CMZ. Accretion of gas onto the CMZ is obtained automatically as a result of the simulation.

A hydrodynamic simulation such as the one described above can become unstable when one reaches a sufficiently high resolution. This is commonly observed in simulations of this type \citep[e.g.][SBM15a]{Kim++2012a}, and is believed to be a real physical phenomena, dubbed the ``wiggle instability'' \citep{WadaKoda2004,KimKimKim2014}. We will discuss this further in Section \ref{sec:instability}. Nonetheless, averaged over time the large-scale features of the flow remain approximately constant.

\subsection{Nuclear spirals}

\begin{figure}
\includegraphics[width=0.5\textwidth]{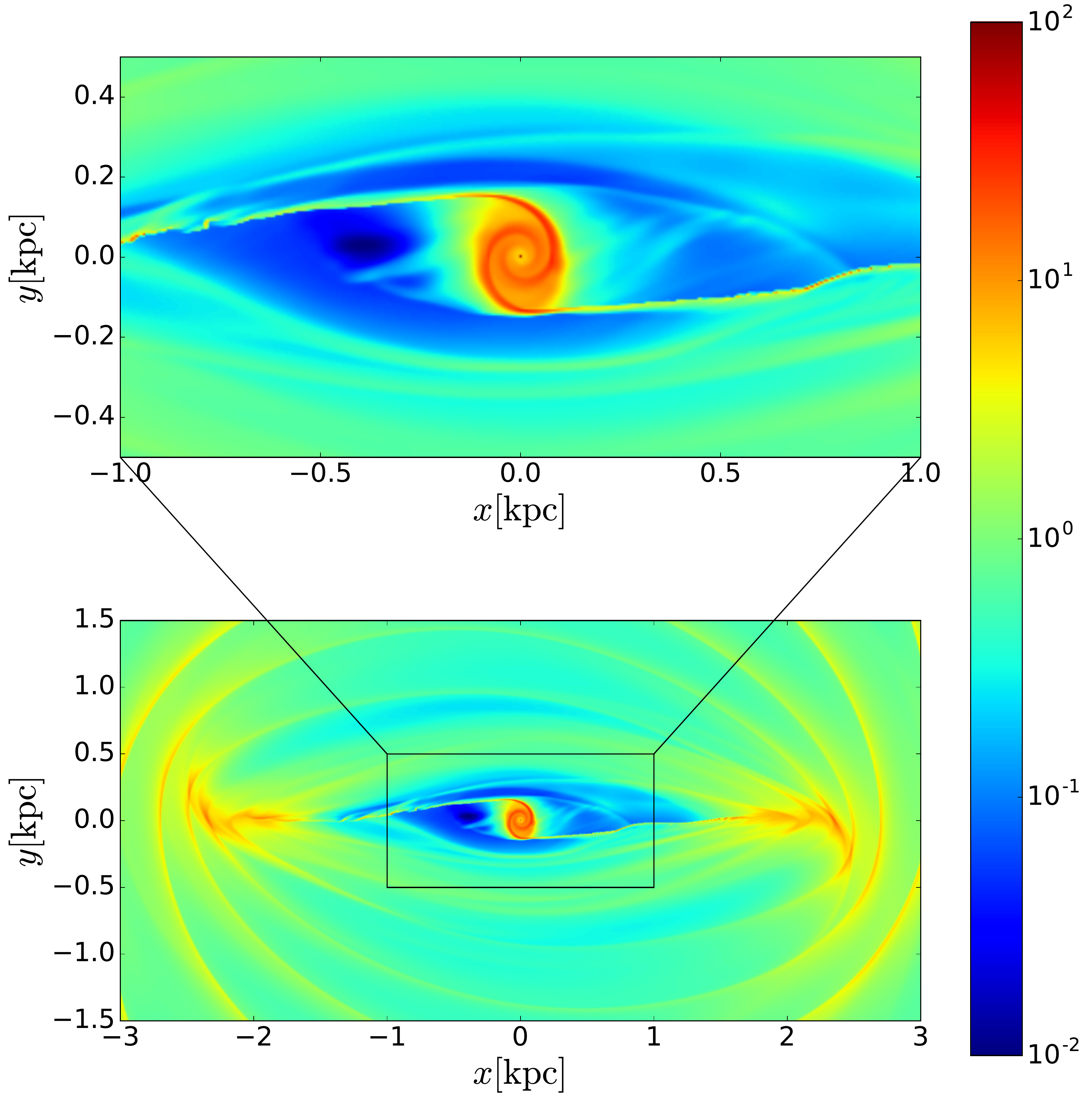}
\caption{Bottom: the fluid density of the simulation at 280Myr, after the flow has reached an approximate steady state. Top: a closer view of the central region of the simulation. The colorbar is in units of $M_\odot \pc^{-2}$.}
\label{fig:rho}
\end{figure}

Nuclear spirals are evident in the inner region ($R\lesssim 500\pc$) in Fig. \ref{fig:rho}, with the outer end of each spiral arm connected to the inner end of each shock. We propose below that some of the streams observed in the data are associated with such nuclear spirals. The presence of nuclear spirals is not in contradiction with the fact that gas approximately follows $x_2$ orbits. Indeed, the gas does follow $x_2$ orbits well, but not \emph{exactly}: tiny \emph{librations} around the $x_2$ orbits ``interfere constructively'' and generate the spiral arms as kinematic density waves, by exactly the same mechanism described in \cite{SBM2015b}, see their figs 10, 11 and 12. The only difference is that here librations are around $x_2$ orbits rather than $x_1$ orbits as in \cite{SBM2015b}.

It follows from the nature of the spiral arms as kinematic density waves that gas does not flow \emph{along} the spiral, but has a significant component of the velocity perpendicular to it. In other words, streamlines cross the spiral arms at an angle, and a fluid element that is instantaneously on a spiral arm moves away from it at a later time. In our simulation, gas usually gets compressed when entering the spiral arm and decompressed when leaving it again. However, in the real ISM if material is compressed in the spiral arm to the point that a cloudlet that can be considered as a separate entity is formed, this cloudlet will continue to move along the streamline, leaving the spiral arm. Therefore it is possible for an overdense cloudlet to leave the spiral, despite larger-scale decompressions, and remain recognisable as an individual entity. Even in our ``smooth'' simulations, we have observed this behaviour: material flowing down the shock and entering the inner spiral can fragment as a consequence of the wiggle instability (see Section \ref{sec:instability}). These fragments are observed to move out from the spiral, in the manner that droplets of water leave the surface of a spinning football. These fragments/cloudlets will continue on approximate ballistic orbits around the CMZ, until colliding with material that has been falling down the shock near the point where it feeds the other spiral arm.

The precise morphology of the nuclear disc and spiral arms is affected by the resolution and sound speed of the simulation (SBM15a). This dependency is complex. Both the resolution and the sound speed affect the location at which the shocks form, and hence the location at which gas is fed onto the CMZ.  They also affect the dynamics within the CMZ, such as the stability of the flow (see Section \ref{sec:instability}). We leave a full discussion to later work, and set the resolution (5pc, on the order of the size of a single molecular cloud) and the sound speed to reasonable values in our model in order to focus on discussing how nuclear spirals, if indeed present in the Milky Way, would appear in observations and affect the dynamics of gas within the CMZ.

\section{Discussion}
\label{sec:discuss}

\subsection{Interpretation of $(l,v)$ features and Face-On Map of the CMZ}
\label{sec:features}

Using our model of the gas flow we can understand the features previously identified in the data. Figure \ref{fig:regions} shows a close up view of the central region of the simulation, with several features of the flow highlighted. In red and blue are the outer parts of the near-side and far-side spiral arms respectively, and in green is the positive-latitude shock connecting the tip of the cusped $x_1$ orbit to the $x_2$ disc. The bottom panel shows the projection of these features in the $(l,v)$ plane. Also shown is the NH$_3$ data for comparison.

\subsubsection{Arms \textsc{i} and \textsc{ii}}

In the longitude-velocity plane, the two spiral arms produce two parallel ridges of emission reaching from positive latitude and velocity to negative latitude and velocity (red and blue in the bottom panel of Fig. \ref{fig:regions}). Moving outwards-in along the red (blue) arm in the top panel of Fig. \ref{fig:regions}, the corresponding trace in the bottom panel is from negative (positive) longitude and velocities to positive (negative) longitude and velocities. As the tangent point between the line of sight and the spiral arm is approached, approximately where the red and blue colours end in Fig. \ref{fig:regions}, the direction of movement in the $(l,v)$ plane is reversed and we go back again towards negative (positive) longitudes.

The two parallel ridges in the bottom panel of Fig. \ref{fig:regions} have a similar range in longitude and a similar slope to Arm \textsc{i} and \textsc{ii} in Fig. \ref{fig:features}. Thus we identify Arm \textsc{ii} (\textsc{i}) as the projections of the near (far) side nuclear spiral arm. These two features were identified with two spiral arms also in the interpretation of \cite{Sofue95}. However, in the interpretation of this author the two arms were swapped in the $(l,v)$ plane: they identified the Arm \textsc{i} (\textsc{ii}) as the projections of the near (far) side nuclear spiral arm, the opposite of our interpretation. This difference allows us to correct an inconsistency of the spiral arms of \cite{Sofue95} pointed out by \cite{Henshaw+16a} regarding the placements on the $20\kms$ and $50\kms$ clouds (see Section \ref{sec:previouswork} below).

We interpret Arm \textsc{iii} (the "Polar Arc") as a spur of gas extending from Arm \textsc{ii} (as proposed in \citealt{Binney++1991}), however we note that our simplified 2D model cannot explain its position at high inclination.

\subsubsection{1.3$\degree$ cloud complexes} \label{sec:B2}

Figure \ref{fig:faceon} shows a face-on view of the CMZ, together with streamlines of the gas flow. As discussed in Section \ref{sec:results}, streamlines do not follow the spiral arms, but go through it at an angle. A dense cloudlet, formed for example as a consequence of compression in the arm, can leave the spiral arm and continue its course along a streamline. Such a cloudlet eventually collides with material that has been flowing down the positive-latitude shock, at a position that depends on where it is ejected from the spiral arm. Such collisions would create cloud complexes with complicated spatial and velocity structure.

We have observed this happening in our simulations: the gas fragments due to the wiggle instability, and the fragments tend to leave the spiral arms by following the streamlines and collide with material entering the other arm. Similar behaviour is seen, e.g. in the simulations of \cite{Dobbs2008} or \cite{Smith2014} who model of the dynamics of the ISM and molecular cloud formation during spiral arm passage on large galactic scales. In the real ISM, a cloudlet could form also by other means, for example as a consequence of becoming quasi self-gravitating after compression. However, we do not include self-gravity in the present paper.

We also expect material to be ejected out of the plane of the CMZ as a consequence of collisions. This could explain why the observations look thicker in latitude on the left and right edges of the CMZ (top panel of Fig. \ref{fig:features}), where the areas surrounding Sgr B2 and 1.3$\degree$ cloud complexes on one side and Sgr C on the other side are suggestive of two ``lobes'' at the sides of the CMZ. Thus the blue and pink on one side and the grey on the other side in the top panel of Fig. \ref{fig:features} could be the two sites where the shocks feed the two inner spirals.

Thus we interpret the 1.3$^\circ$ cloud complex as the result of cloud-cloud collisions between the end of the dust-lane shock and the inner disc, and Sgr B2 as material that has detached from the red arm (discussed further below). In this interpretation, the observed complicated velocity structures and unusually large vertical extent of the gas (see cyan and pink in the top panel of Fig. \ref{fig:features}) are created by collisions (as also interpreted by \citealt{Bally+10}).

\subsubsection{Placement of prominent molecular clouds}

An understanding of the history of the environment of a star forming molecular cloud would give insight into the physical processes that can trigger and regulate star formation. Several gas clouds in the CMZ have been identified as the possible progenitors of star clusters \citep{Longmore+2013b}, the locations of these prominent cloud complexes within the CMZ are therefore of particular interest because they represent an excellent opportunity to study star formation in extreme conditions \citep{Kruijssen+14b}.

Using the steady-state velocity field of the gas flow any $(l,v)$ point can be deprojected to one or more $(x,y)$ positions: in practice there will often be two or more points along a line of sight with the same line of sight velocity, so additional information is required to map a given point in $(l,v)$ space to the $(x,y)$ plane.

The Brick \citep{Lis1998,Longmore+2012,Rathborne+2015} and the 20$\kms$ and 50$\kms$ clouds \citep{Molinari+2011} have been detected in absorption at IR wavelengths, which suggests that they lie in front of the GC. In addition, \cite{Reid2009} measured both the parallax and proper motion of water masers in Sgr B2, placing it at $130\pm6$pc in front of the GC with $(\mu_l, \mu_b) = (2.3\pm1.0, -1.4\pm1.0)\,\rm{mas}\,\rm{yr}^{-1}$. These data break the degeneracy of our model, allowing us to place these clouds within the CMZ.

Plotted in Figure \ref{fig:faceon} are the position of some prominent molecular clouds in our interpretation. The 20 and 50$\kms$ clouds, with Sgr C, lie along the near side spiral arm, in front of the GC.

As noted by \cite{Kruijssen+2015}, the Sgr B2, the dust ridge, and the Brick seem to form a continuous structure in $(l, b, v)$ space. This region of emission (the green points in Fig. \ref{fig:features}) is complex, with multiple components, and it is tempting to identify this structure as the continuation of Arm \textsc{i}. However, this would place the clouds behind the GC, while the data suggest the opposite. Therefore we interpret this region of emission as the superposition of two distinct physical regions in the Galaxy that happen to be at the same location in the $(l,v)$ projections: Arm \textsc{i} behind the GC, and a spur of material detaching from Arm \textsc{ii} on the near side, containing the Brick, dust ridge clouds, and Sgr B2.

This is also consistent with the observations that Sgr B2 is host to a high rate of star formation. The compression created by cloud cloud collisions would provide a natural mechanisms for triggering gravitational collapse and star formation.

We should always take these interpretations with a grain of salt. For example, it is also tempting to connect Sgr B2 and the Brick with Arm \textsc{i}, the far side arm. The presence of the Brick in absorption together with the observed parallax and proper motion of Sgr B2 would go against this interpretation. However, this relies on the assumption that we know the infrared emission patterns within the CMZ, and clouds in front of the GC show absorption and clouds behind the GC are obscured. We should bear in mind that while these assumptions are plausible there is always the possibility that they are incorrect.

\subsubsection{Summary}

To summarise, we interpret:
\begin{enumerate}
\item Arm \textsc{ii} (\textsc{i}) as the projections of the near (far) side nuclear spiral arms.
\item Sgr B2, the dust ridge, and the Brick cloud as a spur of material detaching from the red arm, the end of which (Sgr B2) is colliding with shocked material reaching the inner disc.
\item The 1.3$^\circ$ cloud complex on one side and the area surrounding Sgr C on the other side as the two sites where the shocks feed the two spiral arms.
\item The polar arm as a spur of material detaching from the red arm (at a further out location than Sgr B2).
\item The 20$\kms$, 50$\kms$ and Sgr C clouds as condensations along the near side spiral arm.
\end{enumerate}

We remind the reader that the model outlined here is to be seen more as a ``cartoon'' sketch that provides a framework for interpreting the gas flow in the CMZ rather than a detailed fit to all individual observational features. Our interpretation is the result of simulations that were intended to describe the larger-scale flow pattern in the Galactic bar that were not fine-tuned to model the CMZ. So there is certainly room for improvement. We hope that future 3D models that keep track of the chemical evolution of gas flowing in the CMZ will provide further insight.

\begin{figure}
\includegraphics[width=84mm]{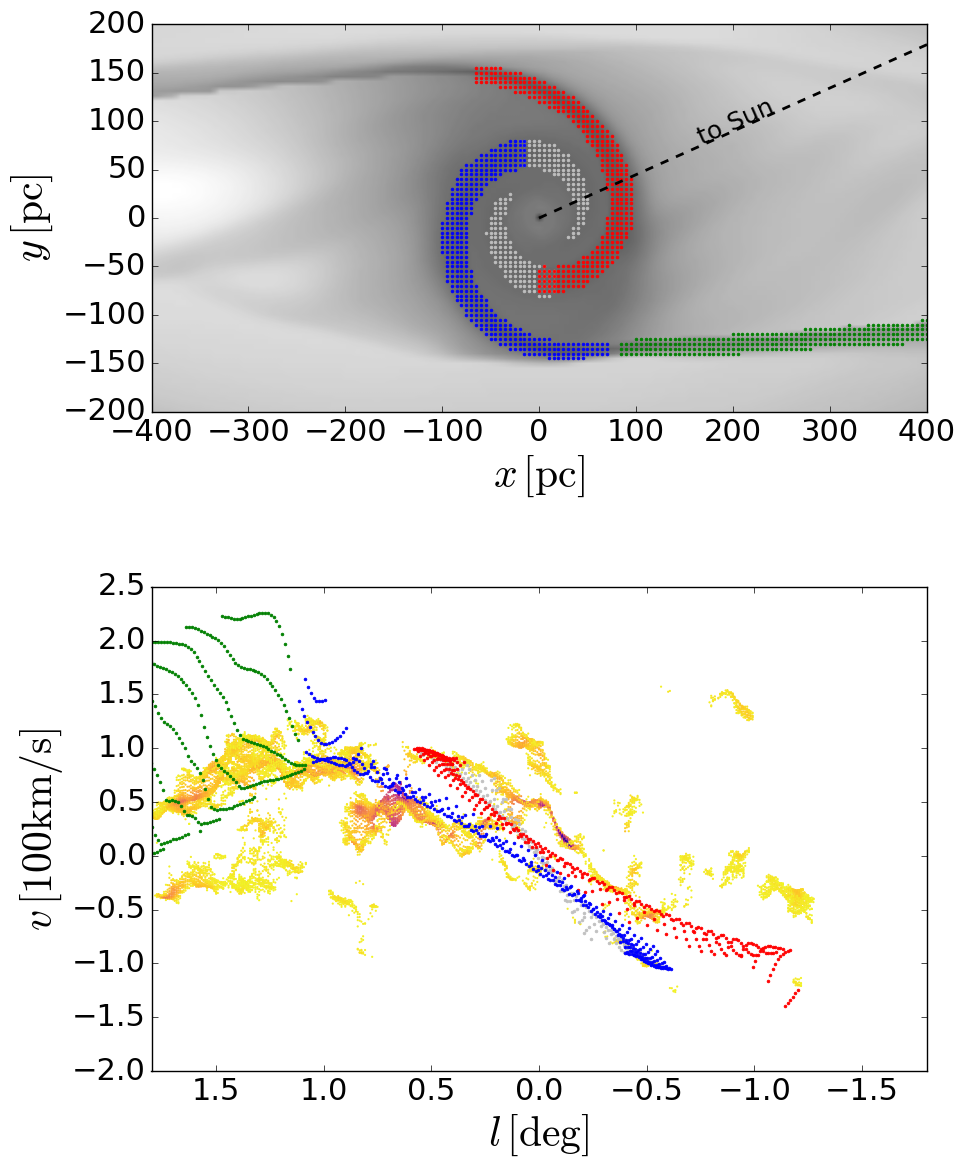}
\caption{Top panel: The near and far side nuclear spiral arms and the positive latitude shock are highlighted in red, blue and green respectively. The inner regions of both arms are also highlighted in gray. Dashed line: The Sun-GC line for our assumed viewing angle of $\phi = 20^\circ$. Bottom panel: Projection of the highlighted features to the $(l,v)$ plane, plotted with the data of Fig. \ref{fig:henshaw}.}
\label{fig:regions}
\end{figure}

\begin{figure}
\includegraphics[width=84mm]{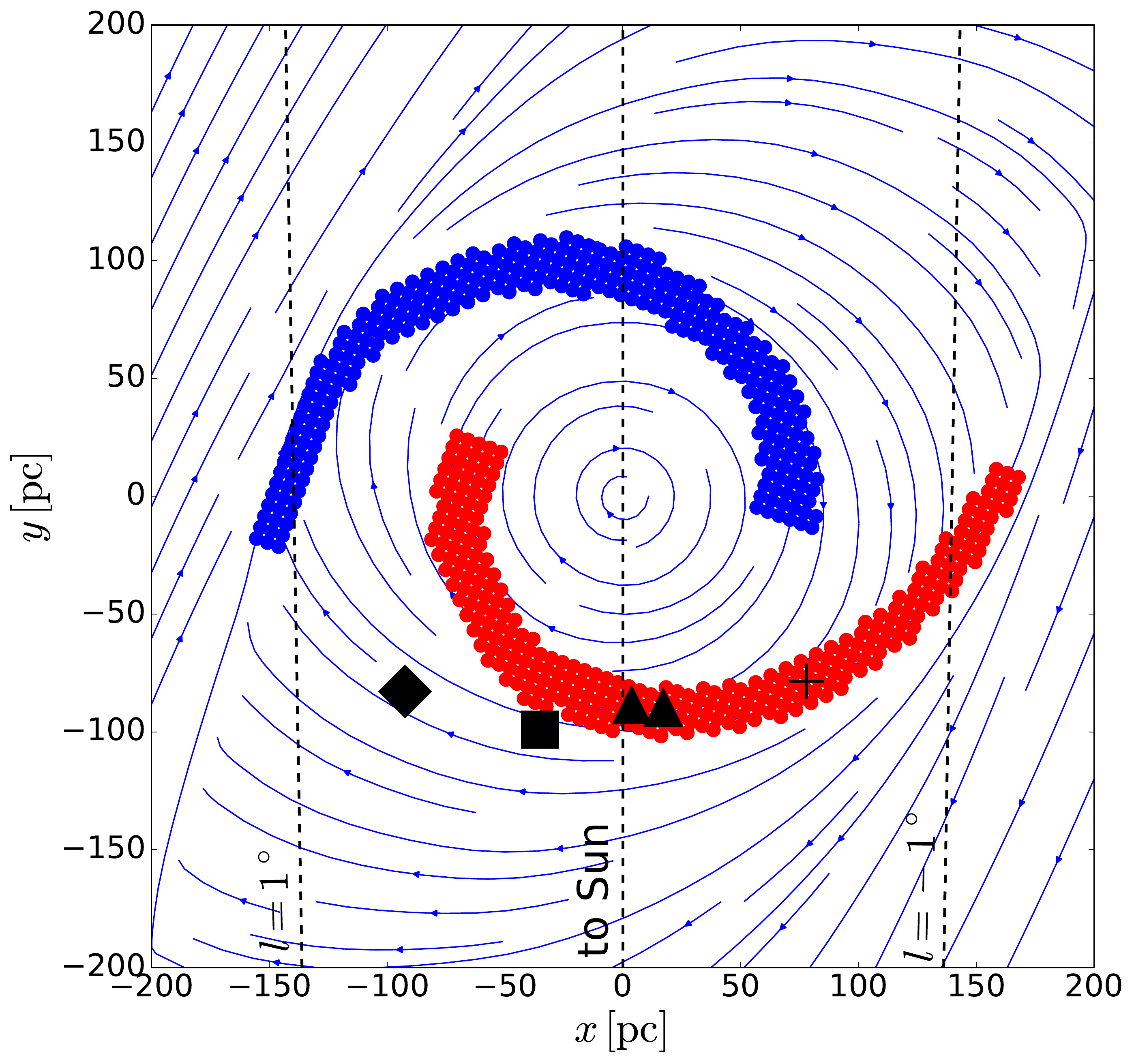}
\caption{A face-on schematic view of the model of the CMZ. The figure has been rotated so that the Sun is at $(x,y) = (0, -8)$. The near and far side spiral arms are shown in red and blue respectively. Streamlines of the gas flow are shown. The black symbols denote the locations of prominent molecular clouds, as in Fig. \ref{fig:features}. From left to right: Sgr B2 (rotated square), G0.253+0.016, also known as ``the brick'' (square), the $20\kms$ and $50\kms$ clouds (upward triangles), Sgr C (plus).}
\label{fig:faceon}
\end{figure}

\subsection{Signatures of unsteady flow}
\label{sec:instability}

\begin{figure}
\includegraphics[width=84mm]{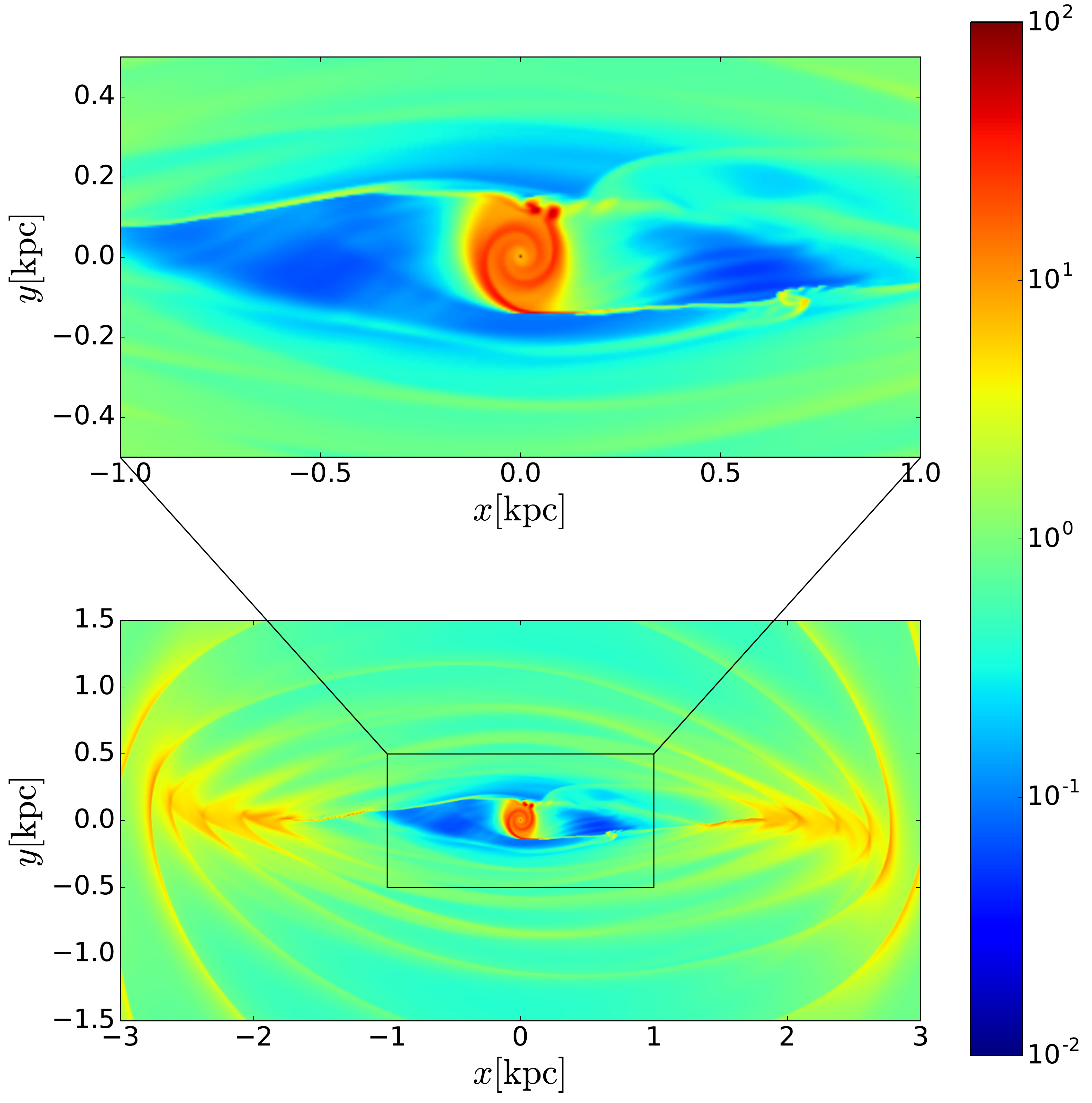}
\caption{Bottom: The fluid density of the simulation at the earlier point of 207$\,$Myr. The flow has reached an approximate steady state, but shows several signs of unsteadiness. Top: A closer view of the central region of the simulation. The colorbar is in units of $M_\odot \pc^{-2}$.}
\label{fig:unstable_rho}
\end{figure}

\begin{figure}
\includegraphics[width=84mm]{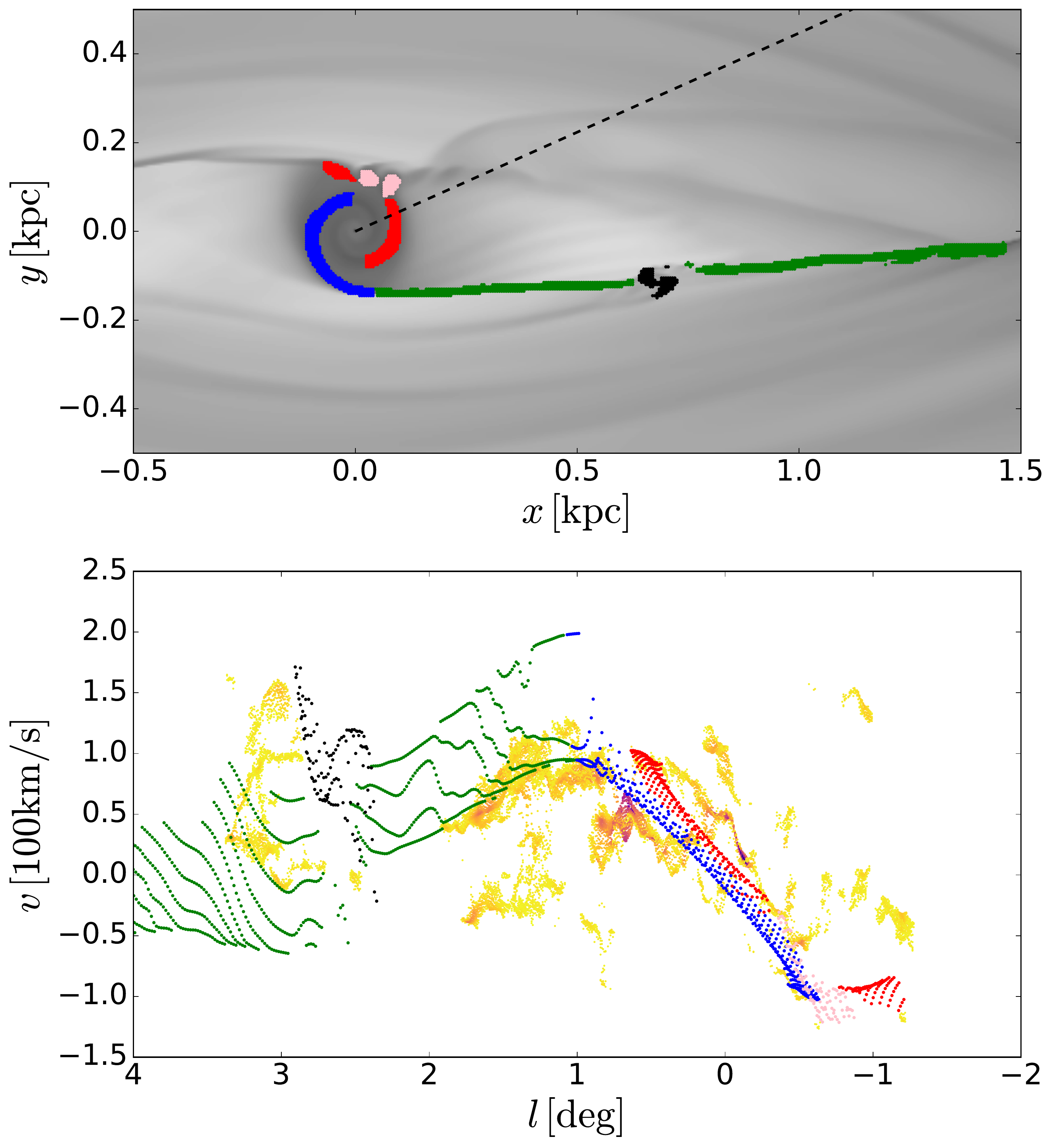}
\caption{The same as Fig. \ref{fig:regions}, for the earlier point of the simulation with two additional features of the flow highlighted: Black points: a ``bead'' of material on the positive latitude shock. Pink points: Two clumps of material along the near-side spiral arm. Both are signatures of the wiggle instability.}
\label{fig:instability}
\end{figure}

The gas flow shows a small amount of unsteadiness. This type of instability, where the shocks continuously develop vortices along their length, break apart, and reform is commonly observed in simulations of gas flow in barred potentials \citep[e.g.][SBM15a]{Kim++2012a,Fragkoudi+2017}, and has been dubbed the ``wiggle'' instability. It does not appear to be a numerical artefact, as it appears in simulations run with many different codes (SBM15a). \cite{KimKimKim2014} used a shearing box analysis to show that the instability originates from the generation of potential vorticity behind the curved shocks. This potential vorticity is amplified as it passes through successive shocks, leading to an instability that periodically destroys the spiral shocks.

Our model is mostly stable, however, there are a few signatures of the wiggle instability in the flow. Figure \ref{fig:instability} shows the gas density at $t=207\Myr$ in the simulation. The flow in the inner region shows some signs of unsteadiness. Most noticeably, a bead of material has formed along the positive latitude shock, and two clumps of material have formed along the near-side spiral arm. The features are highlighted (in black and pink respectively), in Figure \ref{fig:instability}, along with their counterparts in the $(l,v)$ plane.

The two pink ``clumps'' are formed as a consequence of the wiggle instability. Dense material belonging to the pink region is seen detaching from the spiral arms and later colliding with material on the blue arm, close to the point where it is fed by the green shock, as in our interpretation of the Sgr B2 and 1.3$^\circ$ cloud complexes in Section \ref{sec:B2}.

\subsubsection{Clump 2 cloud complex}

It is particularly tempting to associate the signature of the black ``bead'' along the shock in Figure \ref{fig:instability} with the Clump 2 cloud complex. They are almost coincident in $(l,v)$ space, and have a similar velocity structure. We therefore interpret the Clump 2 cloud complex as material that is in the process of transitioning from the $x_1$ to the $x_2$ orbits, part way down the shock, as originally suggested by \cite{StarkBania1986}.

The rest of the emission from the shock (in green) covers a broad region of positive velocities with $l \geq 1.5^\circ$. Therefore it is natural to ask why we do not see emission covering this region in the MW. In practice, if one performs a simple radiative transfer calculation (assuming material is optically thin) the two shocks are almost invisible in projection (see for example figure 5 in \citealt{SBM2015c}), because a limited amount of fast moving material is spread over a large area in the $(l,v)$ plane.

The actual distribution of emission from the shocks will depend on the positions of clouds along the shock and the efficiency of conversion from atomic to molecular gas (SMB15a), and both will be affected by the wiggle instability. As such, it is unsurprising that only a small portion of the $(l,v)$ region covered by the shock is visible in emission. \cite{SBM2015c} has suggested that all the vertical features identified in the $(l,v)$ plane (which include the Clump 2 complex, see their figure 3) are different portions of the two shocks. Simulations including live chemical networks to test this hypothesis are in preparation.

\subsubsection{Asymmetry and star formation}

The wiggle instability might be responsible for at least three other important facts related to the CMZ.

\begin{enumerate}
    \item Approximately three-quarters of the molecular emission from $|l|\la4\degree$ comes from positive longitudes \cite[e.g.][]{Bally1988,Burton+1992}. In particular, we note that Arm \textsc{ii} is far weaker in emission than Arm \textsc{i}, and there appears to be no counterpart to the Sgr B2 and 1.3$^\circ$ cloud complexes at negative latitudes. The cause of this asymmetry is a long-standing puzzle. The asymmetry is too big to be attributed solely to a perspective effect from an inclined bar \citep{JenkinsBinney94}. SBM15a have argued that, at least in part, the marked asymmetry of molecular emission in the Galactic Centre can be explained by unsteady flow. Wiggle instabilities in the shocked flow feeding the CMZ will give rise to unsteady conversion of atomic to molecular gas, so the atomic/molecular ratio on each side of the Centre would fluctuate widely. If the conditions in the GC are suitable to produce the wiggle instability, the nuclear spirals will continuously break apart and reform, and the flow of material down the shocks will be intermittent. This provides a natural explanation for the asymmetry: at the current point in time the near side spiral arm is in a more-intact state than the far side, and the flow of material down the negative latitude shock is in a quiescient phase opposed to the positive latitude shock, which is actively feeding the nuclear disc.
    \item The star formation rate in the CMZ is lower by a factor of $\simeq 10$ than expected from current theories of star formation. \cite{Kruijssen+14b} argue that turbulence is the likely cause for this anomaly \citep[see also][]{Bertram+2015}. The wiggle instability may also be an important source of turbulence in the CMZ.
    \item \cite{Henshaw+16b} recently reported the detection of ``corrugations'' in a stream of gas within the CMZ, which they identify as gas streaming towards molecular cloud condensations seeded by gravitational instabilities. However, if the spiral shocks are being deformed by the wiggle instability, this would also produce an oscillation in the $(l,v)$ structure very similar to observations.
\end{enumerate}

\subsection{3D distribution of gas}
\label{sec:3D}

Molecular emission from the Galactic Centre has a non-trivial vertical structure, as can be also seen in the top panel of Fig. \ref{fig:henshaw}. In a series of papers, \cite{BurtonLiszt1978,LisztBurton1978} and \cite{lisztburton1980} have argued that the inner region of the Galaxy is tilted, so that gas in the central disc ($R\lesssim {\rm few} \kpc$) does not lie in the plane $b=0$ (see also \citealt{Heiligman1987,Ferriere2007} and section 9.4 of \citealt{BM}). The main observational evidence for the tilt is:
\begin{enumerate}
\item In the central disc, emission at positive (negative) longitudes mostly lies at negative (positive) latitudes.
\item Longitude-velocity diagrams obtained by slicing the HI and CO data cubes along a line parallel to $b=-l \tan 22\degree$ rather than $b=0$ show a much higher degree of symmetry.
\end{enumerate}

Although our model is two dimensional, we can produce crude $(l,b)$ maps by modelling the CMZ as a tilted razor thin disc. The orientation of the normal $\hat{\bf{n}}$ to the disc can be specified by two angles $(i,\alpha)$, which are usual polar angles of a spherical coordinate system whose zenith lies along the GC-Sun line and correspond to the angles with the same names in \cite{BurtonLiszt1978}.\footnote{If $XYZ$ are cartesian right-handed coordinates centred at the GC with the Sun lying on the positive $Z$ axis and $X$ pointing towards the north Galactic pole, then
\begin{equation} \hat{\bf{n}} =  \sin i \cos\alpha \hatX +  \sin i \sin \alpha \hatY + \cos i \hatZ \end{equation}}
$i$~corresponds to the usual inclination angle such that $i~=~0$ and $i~=~90\degree$ respectively represent face-on and edge-on discs, while $\alpha$ is the angle that the normal to the disc makes with the line $l=0$ in the plane of the Sky, measured counterclockwise.

\begin{figure}
\includegraphics[width=90mm]{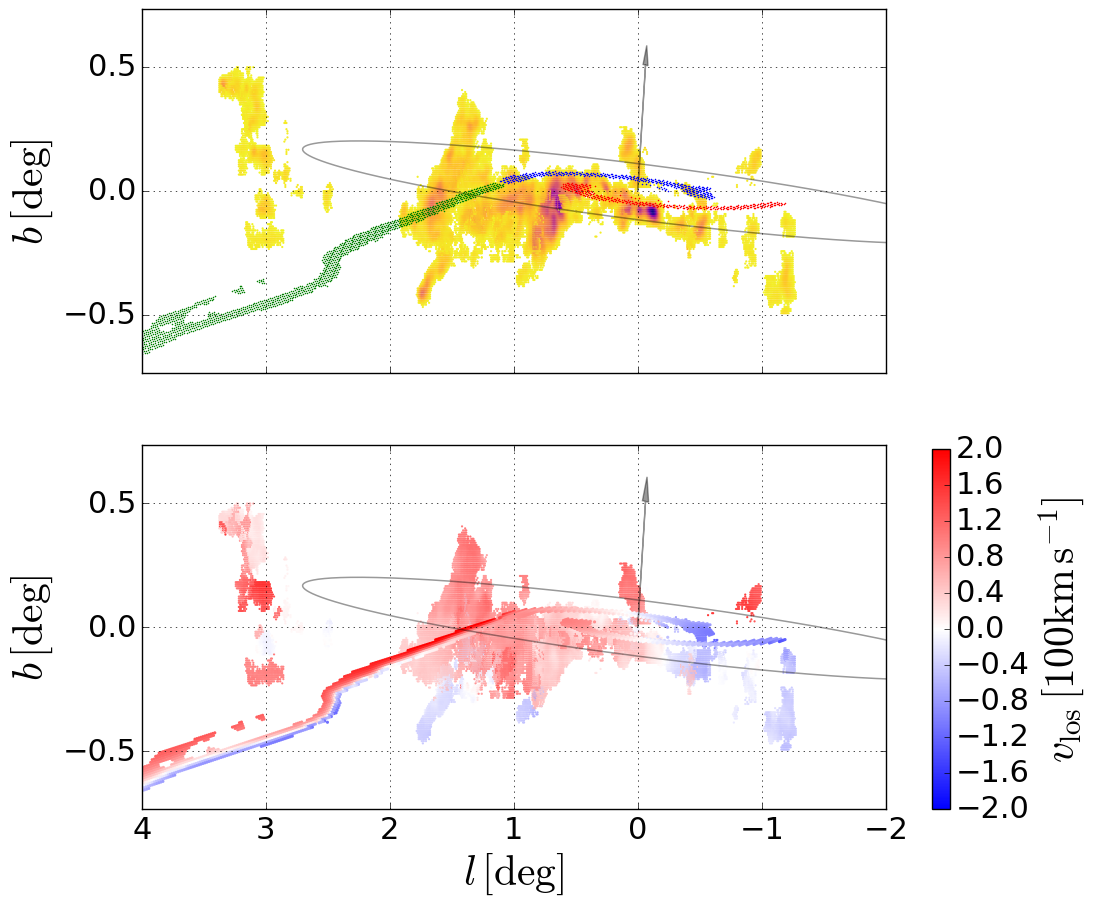}
\caption{A crude 3D model of the CMZ produced by inclining and tilting the simulation as a razor thin disc. The top panel shows the same features highlighted in Fig. \ref{fig:regions}. The bottom panel shows the same, but with both data and model coloured by line-of-sight velocity. The grey arrow shows the normal to the disc $\hatn$, while the grey full line shows the projection of an ellipse elongated perpendicularly to the bar of size $200\times400\pc$.}
\label{fig:tilt}
\end{figure}

Figure \ref{fig:tilt} shows the results of this procedure for $(i,\alpha)~=~(85.7,-2.5)\degree$, which corresponds to an angle of $\theta=5\degree$ between the normal to the disc and the normal to plane of the Galaxy at large. The top panel shows the regions discussed in Section \ref{sec:features} and shown in Fig. \ref{fig:regions}, using the same color code. The bottom panel shows the same, but coloured by line-of-sight velocity. The angle between the GC-Sun line and the major axis of the bar is $\phi=20\degree$ as before.
\begin{enumerate}
\item Within the region $-1^\circ \leq l \leq 1^\circ$ the majority of the emission lines up well along our spiral arms (blue and red). Thus the tilted model captures the $(l,b)$ structure of the streams.
\item Outside the inner degree the gas has a far larger vertical extent than would be expected from our simple model. As discussed in Section \ref{sec:features}, we interpret this region (i.e., the Sgr B2 and 1.3$^\circ$ cloud, green and purple in Fig. \ref{fig:regions}) as the site of cloud-cloud collisions between material at the inner-most tip of the shock and the nuclear disc. These may result in material being thrown out of the plane, producing the kind of structure seen in Fig. \ref{fig:henshaw}.
\end{enumerate}

In our model, $i<90\degree$, which means that the near arm (red) lies at $b<0$, while the far arm (blue) lies at $b>0$. Our value of $i$ is in agreement with previous estimates given in \cite{BurtonLiszt1978,LisztBurton1978} and \cite{lisztburton1980}.

We have found that models with small negative $\alpha$ give a qualitatively better fit to the streams, while previous works favoured $\alpha>0$. Indeed, the first observational evidence mentioned above, i.e. the fact that material at positive (negative) latitude is observed at positive (negative) longitudes, seems to naively imply that $\alpha>0$. However, Fig. \ref{fig:tilt} shows that there is another possibility: if most of the emission comes from material in and around the green shock then one could still obtain the observed apparent tilt even with a small $\alpha<0$. Moreover, the transition point where gas transits from $x_1$ to $x_2$ orbits, which is likely to be a locus of bright emission, also lies in the quadrants that give the correct observed tilt in the $(l,b)$ plane. This reconciles our model with previous estimates. Of course, this is not the only possible explanation. We have assumed that the whole inner region lies on the same plane, but it is also entirely possible that the orientation of the disc changes with radius.

Our discussion is based on purely geometrical considerations. To the best of our knowledge, the dynamical reason for the tilt is at present unknown. Ultimately, real 3D models rather than tilted 2D models are needed to fully understand the vertical structure of the CMZ.

\subsection{Comparison with previous work} \label{sec:previouswork}

The presence of spiral arms at the centre of the MW has been discussed for some time. However, ours is the first dynamical model that takes the larger scale Galactic bar into account consistently, and differs from previous models in several significant ways.

The model of \cite{Sofue95} is purely kinematic. They assumed that gas flows parallel to the spiral arm. As we have already emphasised, the gas in our model does not flow \emph{along} the spiral, but rather \emph{across} it, which allows clouds to leave the spirals. This is also a key difference between our model and the open stream model of \cite{Kruijssen+2015}. In the latter, the observed streams in $(l,v)$ plane are produced by gas flowing along a single open ballistic orbit, and therefore one could try to identify a temporal sequence as the gas evolves along the orbit \citep[an hypotheses suggested by][]{Longmore+2013b}. In our model instead the observed streams are the projection of spiral density waves, i.e. density enhancements, which are caused by librations around $x_2$ orbits as discussed in Section \ref{sec:largerscale}. Whether a temporal sequence could be identified is less straightforward in our model, especially on timescales longer than a fraction of the dynamical time needed to cover an $x_2$ orbit.

Another difference between ours and the model of \cite{Sofue95} is that the two arms are swapped in the $(l,v)$ plane: they identified the near Arm \textsc{ii} (\textsc{i}) as the projections of the near (far) side nuclear spiral arm, the opposite of our interpretation.

The spiral arm models of \cite{Sofue95} and \cite{Sawada+2004} was reconsidered by \cite{Henshaw+16a}, but it was discounted in favour of the open stream model of \cite{Kruijssen+2015}. This was primarily because the configuration of arms that was considered placed the 20$\kms$ and 50$\kms$ clouds on the far side of the GC, and as these clouds are seen in absorption \citep{Molinari+2011} they are most probably on the near side. The model we present here corrects this inconsistency, as the 20$\kms$ and 50$\kms$ clouds are connected to Arm \textsc{i}, which we place on the near side of the CMZ (the red points in Fig. \ref{fig:regions}).

\cite{Kruijssen+2015} have argued that observations suggest that Arm \textsc{ii} and the Sgr B2 and 1.3$^\circ$ cloud complexes are unconnected, separate features (see the red circle in their fig. 2 and the discussion in their section 2.3). In our model this is indeed the case, as they correspond to projections of physically distinct parts of the gas flow. In contrast, they were connected in the model of \cite{Molinari+2011}.

Using an unbarred, axisymmetric potential based on the density profile of \cite{Launhardt+2002}, \cite{Kruijssen+2015} showed that if one assumes that the coherent gas streams in the CMZ all lie along a single ballistic orbit, then an orbit can be found that fits the observed $(l,b,v)$ distributions of the streams well. Their model fails to explain how gas from the larger-scale flow might end up on such an orbit, however. Our model was not originally intended to provide a fit to the CMZ data, but instead was constructed to match the larger scale flow pattern in the Galactic bar. In that sense, it is a "by-product" that it also happens to provide a natural explanation for the origin of the coherant features seen across the CMZ. It automatically accounts for the inflow of shocked gas into the CMZ, and the effect of the non-axisymmetric potential of the bar, both of which are likely to have important consequences on the characteristics of the gas dynamics in the CMZ. We do not assume that the gas streams lie along a single orbit, and indeed find that in our model this is not the case. For example, according to our model, the 20 and 50 $\kms$ clouds are probably not along the same streamline, but on neighbouring distinct streamlines.

The main weakness of our model is that the projections of the spiral arms are at lower velocities than their observational counterparts. In particular, the near-side arm (red in Fig. \ref{fig:features}) would need to be at higher line-of-sight velocities to fully match the observations. However, we note again that our model is not tuned to fit the CMZ. The potential is a relatively simple multi-component model of the Galaxy, in which the quadrupole was chosen to fit several completely different large scale features of the Galactic $(l,v)$ diagram rather than the CMZ \citep{SBM2015c}. It is thus remarkable how closely the model reproduces the majority of the observed features in the CMZ, although features in our simulation and those in the data do not line up perfectly. This also leaves ample space open for improvement: it is very likely that a better fit can be obtained by fine-tuning the potential and/or making the simulations more sophisticated, for example making them 3D, adding self-gravity, or moving on from the simple assumption of an isothermal gas by adding heating and cooling sources and keeping track of the chemistry of the ISM.

There are several other factors which may affect gas flow in the CMZ not considered here, primarily magnetic fields and stellar feedback. \cite{Crocker2010} found a lower limit of 50$\,\mu$G for the magnetic field on $400\pc$ scales near the Galactic Centre, however \cite{Kruijssen+14b} argued that, assuming equipartition, gas densities in the CMZ are above the critical density at which turbulent pressure dominates over magnetic pressure. They also argue that the observed star formation rate in the CMZ is too low for stellar feedback to overcome turbulent pressure. Therefore, following previous work, we assume that the dominating factor driving gas flow on the scales considered here is the response to the gravitational potential.

\section{Conclusion}
\label{sec:conc}

In this paper we have shown that several features present in ($l,b,v$) data cubes of molecular emission from the CMZ can be reproduced by nuclear spiral arms arising from gas flow in a barred potential. We have presented a simple hydrodynamical model of isothermal gas moving in an externally imposed barred potential, which was designed to reproduce features of the gas flow on a much larger scale. In the simulation, a central disc of gas on $x_2$ orbits develops, containing two spiral arms. The disc is connected to the outer regions of the simulation by two shocks.

The model provides a very natural explanation for the structure of the CMZ. Nuclear spirals are common in external galaxies, arise frequently in simulations and are consistent with the larger scale flow in and around the bar.

Although our model is not tuned to the CMZ, it does nevertheless successfully reproduce many aspects of the data. We have shown that the ridges and clouds seen in the data can be understood as the projection of the spiral arms and the shocks to the $(l,v)$ plane. In particular:
\begin{enumerate}
    \item The two spiral arms produce two parallel ridges in the $(l,v)$ plot, running diagonally from positive $(l,v)$ to negative $(l,v)$.
    \item In the region where, in our interpretation, the shock connects to the $x_2$ disc, cloud-cloud collisions are expected, both between material detaching from the spiral arms and between material running down the shock into the nuclear disc. Large cloud complexes with complex line-of-sight velocity structure, such as Sgr B2 and 1.3$^\circ$, are examples of the result of such collisions from material detaching from the red arm and shocked material reaching the inner disc.
    \item It is possible for cloudlets and spurs of material to detach from the spiral arms as a consequence of the fact that streamlines have a component of the velocity perpendicular to the spiral arms. We interpret the polar arm and the dust ridge as an example of such a spur.
    \item A bead of material moving down the positive-latitude shock would produce a vertical emission feature at positive $l$, disconnected from the CMZ, similar to Bania Clump 2.
    \item The compression produced by spiral shocks in the CMZ provides a natural mechanism for producing the densities and pressure necessary for the production of the molecular species that we observe in the region.
     \end{enumerate}

We have found that we can also capture the vertical structure (i.e. in the $(l,b)$ plane) of the emission if we model the CMZ as a razor-thin disc whose axis is tilted by $5 \degree$ with respect to the axis of the Galaxy at large. This fits with previous findings, but the dynamical reasons for the tilt remain unknown.

Moreover, we find that the wiggle instability, often seen in simulations of gas flow in barred potentials, may provide a natural explanation for two important CMZ facts:
\begin{enumerate}
\item The observed asymmetry of emission. Some particular signatures of unsteady flow are present in both the simulation and the data.
\item The low star formation rate observed in the CMZ. The wiggle instability would provide the source of turbulence needed to suppress star formation.
\end{enumerate}

The features of the model presented here are a qualitatively good match to the data, but there are a number of discrepancies, notably the too small line of sight velocities produced by nuclear spirals in the $(l,v)$ plane. The model is by no means a ``best fit'', and has been chosen to show that a simple gas flow model in a realistic potential can provide a useful and physically appealing interpretation of the features observed in the CMZ.

Directions for future work include:
\begin{enumerate}
\item Improving the fit by changing parameters of the underlying potential (e.g. the bar size, pattern speed, ...) or the gas flow (e.g. the sound speed of the gas).
\item Use simulations that include live chemical networks to test the hypotheses regarding i) the origin of the asymmetry and ii) the origin of vertical features such as Bania Clump 2. Such simulations are under way.
\end{enumerate}

\section*{Acknowledgements}

The authors thank James Binney, Simon Glover, Diederik Kruijssen and the referee Mark Wilkinson for insightful comments and suggestions, and Jonathan Henshaw and Steve Longmore for stimulating discussions and for kindly providing the NH$_3$ data. MR and JM acknowledge support from ERC grant number 321067. MCS and RSK acknowledge support from the Deutsche Forschungsgemeinschaft in the Collaborative Research Centre (SFB 881) ``The Milky Way System'' (subprojects B1, B2, and B8) and in the Priority Program SPP 1573 ``Physics of the Interstellar Medium'' (grant numbers KL 1358/18.1, KL 1358/19.2). RSK furthermore thanks the European Research Council for funding in  the ERC Advanced Grant STARLIGHT (project number 339177).

\def\aap{A\&A}\def\aj{AJ}\def\apj{ApJ}\def\mnras{MNRAS}\def\araa{ARA\&A}\def\aapr{Astronomy \&
 Astrophysics Review}\def\apjs{ApJS}\def\apjl{ApJ}\def\pasj{PASJ}\def\nat{Nature}\def\prd{Phys. Rev. D}
\def\ssr{Space Sci. Rev.}\def\pasp{PASP}
\bibliographystyle{mn2e}
\bibliography{bibliography}

\end{document}